\documentclass[prl,amssymb,twocolumn,showpacs]{revtex4}
\usepackage{bm}
\usepackage{graphicx}
\usepackage{amsmath}
\usepackage{color}

\begin{document}

\title{Local density of states in disordered two-dimensional electron gases
at high magnetic~field}

\author{Thierry Champel}
\affiliation{Universit\'{e} Joseph Fourier, Laboratoire de Physique et
Mod\'{e}lisation des Milieux Condens\'{e}s, CNRS, B.P. 166,
25 Avenue des Martyrs, 38042 Grenoble Cedex 9, France}

\author{Serge Florens}
\affiliation{Institut N\'{e}el, CNRS and Universit\'{e} Joseph Fourier,
B.P. 166, 25 Avenue des Martyrs, 38042 Grenoble Cedex 9, France }

\date{\today}

\begin{abstract}
Motivated by high-accuracy scanning tunneling spectroscopy measurements on disordered  two-dimensional electron gases in strong magnetic field,
we present an expression for the local density of states (LDoS) of 
electrons moving in an arbitrary potential smooth on the scale of the 
magnetic length, that can be locally described up to its second derivatives. 
We use a technique based on coherent state Green's functions, allowing us to
treat on an equal footing confining and open quantum systems.
The energy-dependence of the LDoS is found to be universal in terms of local geometric
properties, such as drift velocity and potential curvature. 
We also show that thermal effects are quite important close to saddle points,
leading to an overbroadening of the tunneling trajectories.

\end{abstract}

\maketitle

The most remarkable feature of the Quantum Hall Effect
\cite{vonKlitzing2005} lies perhaps in the fundamental role played by local imperfections
for the successful observation of perfectly quantized Hall conductance plateaus
in two-dimensional electron gases (2DEG) at high perpendicular magnetic field. Yet,
many physical aspects of these systems related to spatial disorder, which are crucial to
understand, e.g., the nature of the quantum Hall transitions, are not easily elucidated
from macroscopic measurements. For this reason, many experimental works were
devoted in recent years towards developing local probes \cite{Ilani},
and have unveiled surprisingly rich interplays of disorder and electronic interactions.
The most straightforward local measurement in interpretative terms is certainly
the local density of states (LDoS) obtained from the scanning tunneling spectroscopy (STS). Such
experiments are however impeded by the fact that 2DEG are usually buried several
hundreds of nanometers below the sample surface, while the characteristic length
scale of the Quantum Hall regime, the so-called magnetic length $l_{B}=\sqrt{\hbar c/|e|B}$,
with $\hbar$ Planck's constant, $c$ the speed of light, $e=-|e|$ the electron charge, and $B$
the magnetic field strength, ranges about 10 nanometers at fields of several teslas.
Only very recently could STS spectra be obtained with an accuracy comparable to $l_B$, thanks to a
special surface treatment of InSb semiconductors \cite{Hashimoto}. This experiment has shown
that localized electronic states of an incompressible Quantum Hall fluid follow closed
semiclassical cyclotron orbits, with a spatial smearing on the scale $l_B$ due to
wave function broadening. Near saddle points of the disordered potential landscape, wider
structures were observed, that were linked to quantum tunneling.

In this Letter, we propose a simple analytical theory for the electronic LDoS in a smooth 
arbitrary potential landscape and strong, yet finite,
perpendicular magnetic field, on the basis of a coherent state Green's function formalism
\cite{Champel1,Champel2}.
As a guide to experimentalists, we provide simple analytical expressions for the
energy-dependent LDoS, and distinguish qualitative spectral features
according to whether thermal or quantum smearing prevails. In the latter case,
different line shapes are obtained for nonzero or vanishing drift velocity.
We also show that thermal broadening is much more effective near saddle points, and may contribute
to the broad structures observed in \cite{Hashimoto}.

Focusing on incompressible regions of Quantum Hall systems, we consider the
single-particle Hamiltonian for a charged particle confined to a two-dimensional
${\bf r}=(x,y)$ plane in the presence of both a perpendicular magnetic field and an
arbitrary potential energy landscape $V$:
\begin{equation}
H=\frac{1}{2 m^{\ast}} \left(-i \hbar {\bm \nabla}_{{\bf r}}-\frac{e}{c}{\bf A}({\bf r}) \right)^{2}+V({\bf r}),
\label{Ham}
\end{equation}
with the vector potential ${\bf A}$ defined by ${\bm \nabla} \times {\bf A}=B \hat{\bf e}_z$
and $m^\ast$ the electron effective mass.
The STS spectra at fixed energy $\varepsilon$ are proportional to the temperature ($T$) broadened
local density of states \cite{note}:
\begin{equation}
\label{rhoSTM}
\rho^\mathrm{STS}(\varepsilon,{\bf r},T) = \frac{-1}{\pi} \mathrm{Im} \int d\omega
\frac{ G({\bf r},{\bf r},\omega)}{ 4 T\cosh^2[(\omega-\varepsilon)/2T]}
\end{equation}
obtained from the retarded Green's function $G({\bf r},{\bf r'},\omega)$ of
Hamiltonian (\ref{Ham}). This quantity can be systematically expanded in a power
series of the type $l_B^{n_x+n_y} \partial^{n_x}_x \partial^{n_y}_y V({\bf r})$,
as was shown in Ref. \cite{Champel2}. This can be done by introducing the Landau
level index $m$ together with a continuous quantum number ${\bf R}=(X,Y)$
defining the ``vortex''-like singularity of the following family of coherent
states:
\begin{equation}
\left<{\bf r}|m,{\bf R}\right>
= \frac{1}{\sqrt{2\pi l_B^2 m!}}
\left(\frac{z-Z}{\sqrt{2}l_B}\right)^m
e^{-\frac{|z|^2+|Z|^2-2Zz^\ast}{4 l_B^2}}
\label{vortex}
\end{equation}
with $z=x+iy$ and $Z=X+iY$.
Modified vortex state Green's functions $h_m({\bf R})$ can then be obtained
\cite{Champel2}, which read for simplicity in the limit of vanishing Landau
level mixing ($\omega_{c}=|e|B/(m^\ast c) \to \infty$, but keeping $l_B$ finite):
\begin{eqnarray}
\label{eq:G}
G({\bf r},{\bf r},\omega) & = & \sum_{m=0}^{+ \infty}
\int \!\! \frac{d^{2} {\bf R}}{2 \pi l_{B}^{2}}
\left| \Phi_{m}({\bf R}-{\bf r}) \right|^{2}
h_m({\bf R}) \\
\left| \Phi_{m}({\bf R}) \right|^{2} & = &
\frac{1}{\pi \, m! l_{B}^{2}}
\frac{\partial^{m}}{\partial s^{m}}
\left.\frac{e^{-A_{s} {\bf R}^{2}/l_B^2}}{1+s}
\right|_{s=0}
\end{eqnarray}
with $A_{s}=(1-s)/(1+s)$.
It can be checked \cite{Champel4} by inspection that, up to second spatial derivatives of the potential,
Dyson equation for the Green's function exactly maps onto the following partial
differential equation
\begin{eqnarray}
\nonumber
1&=&
\left[ \omega+i0^+-E_{m} -V({\bf R})-\frac{2m+1}{4} l_{B}^{2} \Delta_{{\bf R}} V \right]h_{m}({\bf R})\\
 \label{eqquadra}
 &+&\!\!\! \frac{l_{B}^{4}}{8}
\left[
\partial_{Y}^{2} V\partial^{2}_{X}+\partial_{X}^{2} V\partial^{2}_{Y}
- 2 \partial_{X}\partial_{Y}V \partial_{X} \partial_{Y}
\right]h_{m}({\bf R}),
\end{eqnarray}
with $\Delta_{{\bf R}}$ the Laplacian operator.

We give here a brief account of previous
works on high magnetic field calculations, emphasizing similarities and differences
to the present analysis. The popular semiclassical guiding center picture obtained at
$l_{B}\to 0$ keeps a quantum treatment of fast cyclotron motion only, and leads to
eigenenergies that are discrete with respect to the Landau level index $m$ but
continuous with the external potential (a precise mathematical formulation was given
in \cite{Entelis,Champel2}). Our quantum approach naturally contains this semiclassical limit,
as seen by keeping the first line of Eq. (\ref{eqquadra}) only, so that
$h_m({\bf R})$ is indeed given by a simple pole, shifted by a $V({\bf
R})$-dependent term from the energy $E_{m}= (m+1/2)\omega_c$ of isolated Landau
levels. With this result, Green's function $G$ naturally encodes
that the wave functions are translation-invariant Landau states with drift
velocity $c {\bm \nabla}V({\bf r})\times \hat{\bf e}_z/(|e|B)$,
mathematically vindicating an early idea by Trugman \cite{Trugman}.
Capturing quantization/dissipation respectively for closed/open systems asks
however for a full quantum treatment of the guiding center, embodied in the
second line of Eq. (\ref{eqquadra}). The technique of projection onto the lowest
Landau level at finite $l_{B}$ developed in \cite{Girvin} bears in this respect
more similarities to ours. However, the formalism of \cite{Girvin} leads to the
resolution of a one-dimensional Schr\"odinger equation for the wave function, while
the present analysis is based on Green's functions for the overcomplete set of
states $|m,{\bf R}\rangle$.
As a result, an expression such as (\ref{eq:G}) is quite powerful for an {\it arbitrary}
potential $V({\bf r})$ that is locally well described by its Taylor expansion,
as this does not rely on a cumbersome parametrization of the equipotential lines,
as required for a complete set of wave functions. We note that the
path integral formalism developed in \cite{Jain} and also based on the use of the
states (\ref{vortex}) seems to suffer from technical difficulties that were not elucidated,
so that general expressions for observables such as the LDoS, to be discussed in this
Letter, were not obtained to our knowledge. Also, the inclusion of Landau level mixing is there challenging, in contrast to extensions of our Green's function formalism \cite{Champel2}.

Eq. (\ref{eqquadra}) is now easily solved by a mapping onto an ordinary differential equation
thanks to the redefinition
$h_{m}({\bf R})=f_{m}\left[E({\bf R})\right]$, where $E({\bf
R})=V({\bf R})-V({\bf R}_{0})$, with ${\bf R}_0$ an arbitrary reference point for
now:
\begin{equation}
1=\left[(\tilde{\omega}_{m}+i0^+-E)+ \left(\gamma E +\eta\right)\frac{d^{2}}{dE^{2}}+ \gamma \frac{d}{dE}
\right]f_{m}(E)
\label{eqf}
\end{equation}
where
\begin{eqnarray}
\tilde{\omega}_{m} &=&\omega-E_{m}-V({\bf R}_{0})-(m+1/2)\zeta
\label{omegatilde}
\\
\label{gamma}
\gamma & = & \frac{l_{B}^{4}}{4} \mathrm{Det} \left.[H_V]\right|_{{\bf R}={\bf R}_{0}}\\
\label{eta}
\eta &=& \frac{l_{B}^{4}}{8} [({\bm \nabla}V\times\hat{e}_z)\cdot H_V \left.({\bm \nabla}V\times\hat{e}_z)]
\right|_{{\bf R}={\bf R}_{0}}\\
\label{zeta}
\zeta&=&\frac{l_{B}^{2}}{2} \Delta_{{\bf R}} \left. V \right|_{{\bf R}={\bf R}_{0}}.
\end{eqnarray}
The coefficient $\gamma$ is proportional to the determinant of the Hessian matrix
$[H_V]_{ij}=\partial_i\partial_j V$ with $i,j=\{X,Y\}$. Its sign determines the
geometrical nature of the critical points at which the gradient of $V$ vanishes
(a saddle point is characterized by $\gamma <0$, while $\gamma>0$ indicates the presence of a local
extremum).
Physically remarkable is the loose analogy of Eq. (\ref{eqf}) to damped Newtonian dynamics of
equipotential lines with a friction coefficient given by $-\gamma$. This consideration already shows
that dissipation, {\it i.e.}, deviation from a simple pole structure for $h_m({\bf R})$, will thus occur
at the saddle points of $V({\bf r})$.

Differential Eq. (\ref{eqf}) is second order in the derivative with
respect to $E$, but linear in $E$. By doing a Fourier transform
to time $t$, it will become quadratic in $t$ and first order in the derivative
with respect to $t$. Enforcing causality, one finds that the solution
to Eq. (\ref{eqf}) is
\begin{equation}
f_{m}(E)=-i
\int_0^{+\infty} \!\!\!\!\!\!\! dt \,
\frac{e^{-i(E+\eta/\gamma) \tau(t)} }{\cos (\sqrt{\gamma} t )}
\, e^{i(\tilde{\omega}_{m}+i0^+ +\eta/\gamma)t},
\label{intpos}
\end{equation}
with $\tau (t) = (1/\sqrt{\gamma}) \tan\left(\sqrt{\gamma}t\right)$.
The integral in (\ref{intpos}) is defined in the sense of Cauchy principal value at
the points $\sqrt{\gamma} t = (n+1/2)\pi$ for $\gamma>0$. Both cosine and tangent
trigonometric functions are replaced by their hyperbolic counterparts in the
case $\gamma<0$. Eq. (\ref{intpos}) thus remarkably describes both closed and open
quantum systems on an equal footing. When $\gamma>0$, $\tau (t)$
is a periodic function of time, so that $f_{m}(E)$ must display
discrete poles, and one can recover, {\it e.g.}, for a rotationally-invariant
quadratic potential the full solution of the Fock-Darwin problem at small Landau
level mixing \cite{Champel4}. In contrast, for $\gamma<0$, the spectrum remains continuous, as
expected from the solution \cite{Fertig} for the saddle-point potential.
Quite interestingly, the convergence of the integral in (\ref{intpos}) for $\gamma<0$ is not
ensured by the $\exp(-0^+ t)$ cutoff function, but by the kernel
$1/\cosh(\sqrt{-\gamma} t)$, with an energy scale $\sqrt{-\gamma}$
related to the curvature of the potential (see discussion below for the
implications to the LDoS). This leads to the appearance of a finite imaginary part
in the self-energy associated to the equipotential lines, hinting at quantum
dissipation effects (see \cite{Champel4} for further discussion).

\begin{figure}
\includegraphics[scale=0.89]{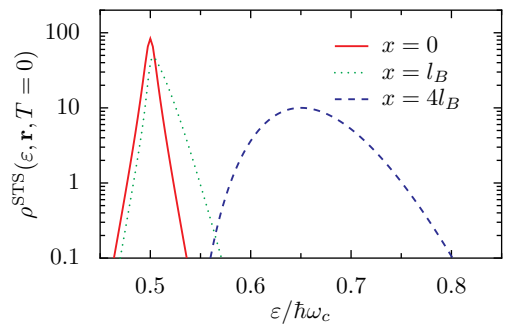}
\caption{(color online) LDoS
from (\ref{rhofinal}) in units of $(2\pi l_{B}^{2})^{-1}$ as a function of $\varepsilon$ at $T=0$ for the saddle-point
potential (\ref{Vsaddle}) with $\omega_0/\omega_c=0.1$.
Here, three different positions were chosen along the bisecting line $x=y$: at the saddle point ($x=0$,
left curve), in the crossover region ($x=l_B$, middle curve), and in the drift
dominated region ($x=4 l_B$, right curve).}
\label{fig1}
\end{figure}

\begin{figure}
\includegraphics[scale=0.89]{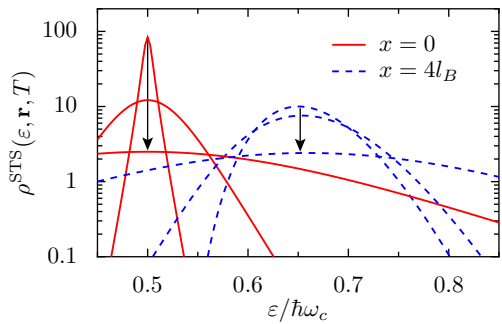}
\caption{(color online) Thermal smearing of the same
LDoS as shown in Fig. \ref{fig1} for the positions
$x=y=0$ and $x=y=4 l_B$ only. Three different increasing temperatures $T/\omega_c=0,0.02,0.1$
were taken (top to bottom, along the arrows).}
\label{fig2}
\end{figure}

Because $h_m({\bf R})=f_m[E({\bf R})]$ obtained in (\ref{intpos}) has a simple exponential
dependence on $E({\bf R})$, it is possible to perform analytically the ${\bf R}$-integration in
Eq. (\ref{eq:G}) by developing $V({\bf R})$ to quadratic order around
${\bf R}_0={\bf r}$. After doing the frequency integral in (\ref{rhoSTM}), the
 LDoS reads
\begin{widetext}
\begin{eqnarray}
\label{rhofinal}
\rho^\mathrm{STS}(\varepsilon,{\bf r},T) = \frac{1}{2 \pi l_{B}^{2}} \mathrm{Re}
\sum_{m=0}^{+\infty} \frac{1}{m!} \frac{\partial^{m}}{\partial s^{m}}
\int_{0}^{+\infty} \!\!\!\!\!\! dt \, \frac{ T t}{\sinh\left[\pi T t \right]}
\left.
\frac{
e^{i[\varepsilon-E_{m}-(m+1/2)\zeta-V({\bf r})]t
+i\frac{\eta}{\gamma}[t-\tau(t)]
-\frac{\tau^{2}(t)}{4} \frac{A_{s} l_{B}^{2} |{\bm \nabla}_{{\bf r}} V|^{2}+4i\eta\tau(t)}
{A_{s}^{2}+i A_{s} \zeta \tau(t)-\gamma\tau^{2}(t)}}
}
{
(1+s) \cos(\sqrt{\gamma}t)
\sqrt{A_{s}^{2}+iA_{s}\zeta \tau(t)-\gamma \tau^{2}(t)}
}
\right|_{s=0}
\end{eqnarray}
\end{widetext}
which is our final theoretical result, to be interpreted at the light of the STS measurements
of Ref. \cite{Hashimoto}.
Although
Eq. (\ref{rhofinal}) is only exact (at small Landau level mixing) for a generic quadratic
external potential (in which case only $\eta$ and $|{\bm \nabla}_{{\bf r}} V|$ depend on ${\bf r}$), it constitutes \cite{Champel4}
a very accurate approximation at finite $T$  for an {\it arbitrary} potential that can be locally described
by the site-dependent geometrical quantities $\{|{\bm \nabla}_{{\bf r}} V|,
\gamma({\bf r}), \eta({\bf r}), \zeta({\bf r})\}$.
Clearly, the function appearing in the integral (\ref{rhofinal}) is governed by
several energy scales, associated to various broadening processes of the LDoS, that we
review in turn (for simplicity, we focus on the lowest Landau
level $m=0$).\\
{\it Quantum smearing from drift motion.} One notices first the appearance of the
energy $\omega_\mathrm{drift}=  l_B |{\bm \nabla}_{{\bf r}} V({\bf r})|$
associated to the drift velocity. When $\omega_\mathrm{drift}$ is the largest energy scale, the smearing of the LDoS is
dominated by the transverse width $l_B$ of the Landau wave functions, and a
good approximation to the LDoS is given by calculating (\ref{rhofinal}) at
$\gamma=\eta=T=0$:
\begin{equation}
\rho^\mathrm{STS}(\varepsilon,{\bf r},0) \approx \frac{1}{2 \pi l_{B}^{2}}\frac{\exp\left[-\left(\frac{\varepsilon-\omega_c/2-V({\bf r})}{\omega_\mathrm{drift}}\right)^{2}\right]}{\sqrt{\pi} \omega_\mathrm{drift}}
.
\label{rhodrift}
\end{equation}
The energy dependence via $\varepsilon$ (controlled experimentally with a back gate
voltage) is thus gaussian in this regime, and is shown in Fig. \ref{fig1}.
We assume for all plots given here the saddle-point potential
profile
\begin{equation}
V({\bf r}) = m^{\ast} \omega_{0}^{2} xy.
\label{Vsaddle}
\end{equation}
We note that small corrections to (\ref{rhodrift}) from the exact formula
(\ref{rhofinal}) provide spectral asymmetries related
to the parameter $\eta$ (see right curve in Fig. \ref{fig1}).\\
{\it Quantum smearing from tunneling at saddle points.} Expression (\ref{rhodrift}) is
certainly invalid at saddle points (${\bf r}=0$) where $\omega_\mathrm{drift}$ vanishes. If the energy
$\omega_\mathrm{saddle}=2\sqrt{-\gamma}$ associated to curvature exceeds both
$\omega_\mathrm{drift}$ and $T$, the LDoS is then better approximated by
performing the integral (\ref{rhofinal}) with $|{\bm \nabla}_{{\bf r}} V| =T=0$
[we take $\zeta=0$ for simplicity, which is satisfied by the potential (\ref{Vsaddle})]:
\begin{eqnarray}
\rho^\mathrm{STS}(\varepsilon,{\bf r},0)  \approx \frac{P_{-1/2+ia}(0)}{2 \pi l_{B}^{2}} 
\frac{\operatorname{sech} \left( \frac{\varepsilon-\omega_c/2-V({\bf r})}{\omega_\mathrm{saddle}/\pi}\right)}
{ \sqrt{2} \omega_\mathrm{saddle}}
\label{rhocurv},
\end{eqnarray}
where $P_{-1/2+ia}(x)$ is the conical function of the first kind \cite{Abramowitz}
and $a=(\varepsilon-\omega_c/2-V({\bf r}))/\omega_\mathrm{saddle}$.
One can deduce from this Eq. (\ref{rhocurv}) that the energy-dependent LDoS has an exponential
behavior near saddle points (see left curve of Fig. \ref{fig1}). {\it Local geometric properties 
of the disorder potential thus have qualitative fingerprints on the LDoS.}\\
{\it Thermal smearing.} When $T$ exceeds both $\omega_\mathrm{drift}$ and
$\omega_\mathrm{saddle}$, the integral (\ref{rhofinal}) is cut by the
$1/\sinh(\pi T t)$ kernel, and is well described by
\begin{eqnarray}
\label{rhothermal}
\rho^\mathrm{STS}(\varepsilon,{\bf r},T) \approx \frac{1}{2 \pi l_{B}^{2}}
\frac{
\operatorname{sech}^{2} \left(\frac{\varepsilon-\omega_c/2-V({\bf r})}{2T}\right)}{4 T}.
\end{eqnarray}
The crossover from the $T=0$ limit to this thermally broadened
form (\ref{rhothermal}) is shown in Fig. \ref{fig2}. These results are easily understood by the fact that
typical values of $\omega_\mathrm{drift}$ far from the saddle point are much larger
than $\omega_\mathrm{saddle}$. For the model potential (\ref{Vsaddle}) one has in particular
$\omega_\mathrm{saddle}=\omega_0^2/\omega_c \ll\omega_\mathrm{drift}=\omega_\mathrm{saddle}
|{\bf r}|/l_B$ for $|{\bf r}|\gg l_B$. This means that {\it thermal smearing will be
more effective near saddle points than onto the drift dominated orbits}, as clearly seen
in Fig. \ref{fig2}.

\begin{figure}[ht]
\includegraphics[scale=0.89]{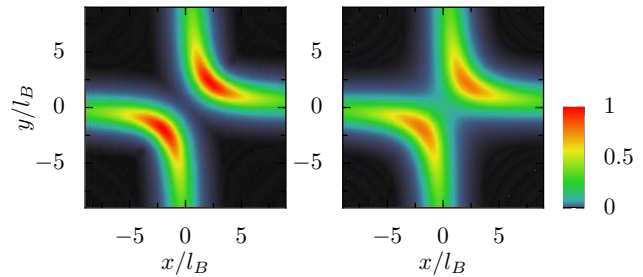}
\caption{(color online) Spatial dependence of $\rho^\mathrm{STS}(\varepsilon,{\bf r},T)$
(normalized to its maximal value at $T=0$) obtained from Eq. (\ref{rhofinal}) for the 
saddle-point potential (\ref{Vsaddle}), with $\omega_0/\omega_c=0.1$,
$\varepsilon/\omega_c=0.54$, at $T=0$ (left panel) and $T/\omega_c=0.01$ (right panel).}
\label{fig3}
\end{figure}

This turns out to be a crucial observation for understanding the spatial
dependence of the LDoS measured by Hashimoto {\it et al.} \cite{Hashimoto}. 
Indeed, our results show that at $T=0$, the LDoS is sharply peaked, 
energetically and spatially, at the saddle points, so that only the tunneling
trajectories that come very close should have appreciable intensity.
However, in the intermediate temperature range $\omega_\mathrm{saddle} \lesssim 
\pi T <\omega_\mathrm{drift}$ [note the $\pi$ prefactor difference in Eqs. (\ref{rhocurv}) 
and (\ref{rhothermal})], important spatial redistribution of the spectral weight
occurs at the saddle-points, 
so that tunneling structures broader than the magnetic length become visible,
see the wide ``bridge'' connecting tunneling trajectories in Fig. \ref{fig3}
at finite temperature.

An important remark concerns how accurate is Eq.
(\ref{rhofinal}) for an {\it arbitrary} potential $V({\bf r})$, since two
assumptions were made in our derivation: (i) large Landau level separation
$\omega_c\gg l_B^2\Delta_{\bf r} V$, {\it i.e.}, $\omega_c \gg \omega_0$ for the
model (\ref{Vsaddle}); (ii) local description of $V({\bf r})$ up to second
order spatial derivatives. Condition (i) is well satisfied in the
experiments, as the analysis of the typical spatial variations of the LDoS
maxima lead to a rough estimate $\omega_0\lesssim3$meV, while $\omega_c=70$meV at
$B=12$T for InSb. We also emphasize that Landau level mixing corrections to all
physical observables can be perturbatively accounted for in our formalism,
as shown in Ref. \cite{Champel2}, so that corrections due to finite
$\omega_c$ are indeed small. In passing, we can also deduce from these experimental 
estimates the energy scale associated to curvature effects
$\omega_0^{2}/\omega_c \lesssim 0.13$meV. Since the base 
temperature $T=0.3$K taken in \cite{Hashimoto} gives a thermal scale 
$\omega_\mathrm{thermal}=\pi T \simeq 0.08$meV, this seems consistent with our 
conclusions on the relevance of thermal smearing near saddle points.
Requirement (ii)
is actually satisfied provided
that temperature exceeds the energy scales associated with third (and beyond) local derivatives of the potential, which turn out to be smaller than the curvature scale $\sim \sqrt{|\gamma|}$.
 We thus conclude that our Eq. (\ref{rhofinal}) yields
quantitative estimates for the LDoS of a weakly disordered electron gas at high
but finite magnetic field.

Other interesting aspects of our calculations, hinted in the text, concern
the question of transport and dissipation in the quantum Hall regime. Although
this topic goes far beyond the scope of the present Letter, we have noted the
natural emergence of quantization and lifetime effects in the respective cases
of closed and open systems from the simple dynamics of equipotentials
lines given by Eq. (\ref{eqf}). Further developments of this idea may be useful in
bringing together a better understanding of local excitations and macroscopic
transport properties of quantum Hall fluids.


\begin{thebibliography}{99}
\bibitem{note} We set now Boltzmann's constant and $\hbar $ to unity, so that
energy, temperature and frequency are equivalently used.
\end{thebibliography}
\end{document}